\documentclass[reprint,showpacs,preprintnumbers,amsmath,amssymb,prl]{revtex4-1}
\usepackage{graphicx,wasysym}
\usepackage{dcolumn}
\usepackage{bm}
\usepackage{natbib}
\usepackage{tabulary}

\newcommand{\bv}[1]{\mathbf{#1}}
\begin{document}

\preprint{1}

\title{Interfacial ordering and accompanying \\divergent capacitance at ionic liquid-metal interfaces}

\author{David T. Limmer}

 \email{dlimmer@princeton.edu}
\affiliation{%
Princeton Center for Theoretical Science, Princeton University, Princeton NJ 08540
}%

\date{\today}
\begin{abstract}
A theory  is constructed for dense ionic solutions near charged planar walls that is valid for strong inter-ionic correlations. This theory predicts a fluctuation-induced, first-order transition and spontaneous charge density ordering at the interface, in the presence of an otherwise disordered bulk solution. The surface ordering is driven by applied voltage and results in an anomalous differential capacitance, in agreement with recent simulation results and consistent with experimental observations of a wide array of systems. Explicit forms for the charge density profile and capacitance are given. The theory is compared with numerical results for the charge frustrated Ising model, which is also found to exhibit a voltage driven first-order  transition.
\end{abstract}


\maketitle

Recently, experimental observations and molecular simulations have suggested a link between long-range structural correlations and the electrochemical response of a double layer capacitor composed of an ionic liquid electrolyte\cite{su2009double,zhong2014resolving,uysal2013structural}. Specifically, these observations have alluded to a possible singular response of the differential capacitance to changes in the applied electric potential\cite{liu2006coexistence}. It has been postulated on the basis of molecular simulations that this response results from a competition between entropic effects of packing and local constraints of electric neutrality within the ionic liquid near a planar, constant potential electrode\cite{merlet2014electric}. Using general arguments, I construct an effective field theory for a symmetric solution of dense ionic media that validates this proposal. This theory explains the observed anomalous capacitance as a result of a first-order interfacial transition associated with spontaneous charge ordering at the electrode surface. 

The interface between a dense ionic solution and a metal electrode has been the subject of much recent study, due to the development of ionic liquid-based supercapacitors that exploit charge separation to create high power energy storage devices\cite{simon2008materials,miller2008electrochemical}. Such concentrated electrolyte solutions exhibit  inter-ionic correlations that render typical mean-field theories developed for dilute solutions, such as Gouy-Chapman-Stern theory\cite{parsons1990electrical}, not applicable. Extensions of these theories to account for excluded volume have been developed\cite{badiali1983microscopic,wu2011classical,bazant2011double,fedorov2014ionic}, 
which are capable of capturing interfacial layering and a nonmonotonic capacitance as a function of applied potential. 
However, such extensions typically assume a linearly responding charge density, which necessitates that their predicted response functions are bounded.  This contrasts molecular dynamics simulations of a model of BMIM$^+$PF$_6^-$ on graphite electrodes that indicates a voltage driven structural transition and divergent capacitance\cite{merlet2014electric}. Experimental indications of similar emergent long-ranged correlations have been observed in many systems, including spontaneous two-dimensional ordering of PF$_6$ on gold\cite{pan20062d} and free surfaces\cite{jeon2012surface}, as well as  observations of hysteresis upon voltage cycling of C$_9$MIM$^+$Tf$_2$N$^-$ on epitaxial graphene \cite{uysal2013structural} with observed structural bistability\cite{uysal2015interfacial}. 

To explain these observations of spontaneous interfacial ordering, I consider the implications of two competing interactions: 1) short range repulsions that arise from packing constraints and can favor spontaneous phase separation, and 2) long-ranged attractions that arise from oppositely charged species and frustrate phase separation. For a symmetric solution, the lowest-order expansion around a uniform charge density yields an effective Hamiltonian,
\begin{eqnarray}
\label{Eq:hamB}
\mathcal{H}_\mathrm{B}[\phi(\bv{r})] = \int_\bv{r}\, && \, \frac{a}{2} \phi^2(\bv{r}) + u \phi^4(\bv{r}) + \frac{m}{2} |\nabla \phi(\bv{r})|^2 \nonumber \\ 
&&+ \frac{Q^2}{2} \int_\bv{r}\,\,  \int_{\bv{r}'}\, \frac{\phi(\bv{r})\phi(\bv{r}')}{|\bv{r}-\bv{r}'|} 
\end{eqnarray}
where, $\phi(\bv{r}) = (\rho_\mathrm{+}(\bv{r}) - \rho_\mathrm{-}(\bv{r}))/\rho$ is local excess charge density, determined by the relative density of cations,  $\rho_\mathrm{+}(\bv{r})$, to anions, $\rho_\mathrm{-}(\bv{r})$, over the mean liquid density, $\rho$. While for asymmetric solutions, a cubic order term is allowed, for this simplistic case, here it is unallowed by symmetry.  The effective coulomb coupling $Q^2 = (z)^2 / \epsilon$, is screened by the optical contribution to the dielectric, $\epsilon$,  and $z$ is the magnitude of the charge of the ions. In principle, the parameters $a,u,$ and $m$ are dependent on temperature and pressure, but here they are taken to all be real constants. The parameters $u$ and $m$ are assumed to be positive, as is necessary to justify the truncation in Eq.~\ref{Eq:hamB}, and $a$ is assumed to be negative and small so that there is an explicit tendency of the uncharged species to demix.  This tendency is supported by observations of spatial clustering in many simple ionic liquids\cite{hu2006heterogeneity}.

Microscopically, the parameters $a,u, m$ and $Q$ are related to the screening and bare correlation lengths. This relation can be clarified by computing the charge susceptibility in momentum space,
\begin{equation}
\label{Eq:Sus}
\chi(\mathbf{k}) = \left ( m k^2 + a + 2\pi Q^2/k^2 \right )^{-1} \, \quad k\ne 0 \, .
\end{equation}
Within the random phase approximation\cite{copestake1982charge,revere1986structure}, the Debye screening length is identified as $\ell_\mathrm{s} =\sqrt{|a|/2\pi Q^2}$, and bare correlation length as $\ell_\mathrm{c}=\sqrt{m/|a|}$ with $u=1$, to set the basic energy scale, or equivalently setting $k_\mathrm{B} T =1$, where $k_\mathrm{B}$ is Boltzmann's constant and $T$ is the temperature\footnote{The small $k$ expansion to Eq.~\ref{Eq:Sus} yields $k^2 \ell_\mathrm{s}^2$ as required from the condition of complete screening.}. For room temperature ionic liquids, bare correlation lengths are expected to be on the order of the size of the molecule\cite{mezger2009layering,mezger2015solid}. For the BMIM$^+$PF$_6^-$ mixture studied in previous simulations, the mean molecular diameter is about 5 $\mathrm{\AA}$\cite{roy2010improved}. Typical Debye screening lengths in ionic liquids and molten salts are between 1 and 2 $\mathrm{\AA}$, due to their low permittivity and large molar volume\cite{wakai2005polar}.
\begin{figure}[t]
\begin{center}
\includegraphics[width=7.cm]{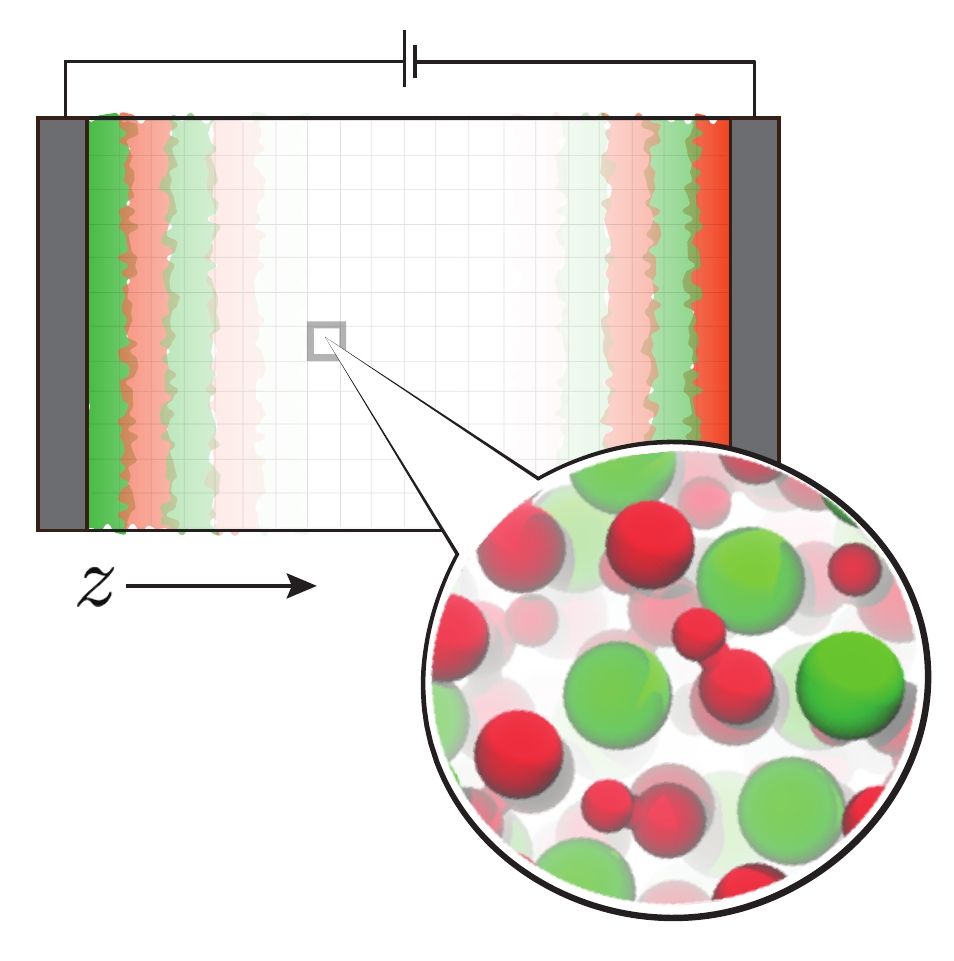}
\caption{Geometry and composition of the capacitive cell considered in this work, namely two ideal parallel plates are separated in the $z$ direction by a solution of nearly symmetric ionic liquid. Under applied potential decaying charge density waves spontaneously form near the interface. The call out contains a snapshot of a typical room temperature ionic liquid taken from a molecular dynamics simulation.}
\label{Fi:1}
\end{center} 
\end{figure}
%
At the Gaussian level of approximation\cite{ortix2008coulomb}, the bulk phase diagram for this model includes a phase transition from a disordered phase for $\ell_\mathrm{s}/\ell_\mathrm{c}<2$, where the $\phi(\mathbf{r}) =0$ on average, to an ordered, microphase-separated state for $\ell_\mathrm{s}/\ell_\mathrm{c}>2$, where $\phi(\mathbf{r})\ne 0$ on average. This microphase-separated state has a characteristic wave vector, $q_\mathrm{s} = 1/\sqrt{\ell_\mathrm{s} \ell_\mathrm{c}}$, as can be deduced from the maximum in the susceptibility in Eq.~\ref{Eq:Sus}, and arises generically from the competition between interactions acting over disparate scales\cite{low1994study}. This specific relationship between the periodicity of charge density oscillations and the correlation and screening lengths has also been arrived at previously\cite{bazant2011double}. In that work however, rather than postulating an effective Hamiltonian for the charge density, Ref. \onlinecite{bazant2011double} derived the expression from a modified Poisson equation. Given typical values of both $\ell_\mathrm{c}$ and $\ell_\mathrm{s}$ for ionic liquids are $\mathcal{O}(1)$, at ambient conditions such liquids are thus expected to be close to microphase separation\footnote{For ionic liquids composed of cations with long hydrocarbon tails, the ratio of $\ell_\mathrm{s}/\ell_\mathrm{c}$ can be very small, placing the bulk in a microphase-separated state\cite{canongia2006nanostructural}.}.  It is the proximity of this phase transition that leads to the anomalous capacitive response, as will be shown below. 

To analyze the interfacial behavior of this theory, a number of simplifications must be made. First, I consider only the case of an ionic liquid in contact with two parallel, identical planar electrodes, a geometry that is illustrated in Fig.~\ref{Fi:1}. In this geometry, the system is symmetric in the plane parallel to the interface, therefore degrees of freedom in the $xy$ plane can be integrated out. Second, the separation between the two electrodes is assumed to be large compared to $\ell_\mathrm{c}$, so that $z$ can be defined over the domain $(0,\infty)$\footnote{This approximation can be relaxed and the order parameter profile solved over a finite domain. See \onlinecite{binder1997surface}}. The resultant effective Hamiltonian per unit area is
\begin{equation}
\label{Eq:hamZ}
\tilde{\mathcal{H}}[\phi(z)] = \tilde{\mathcal{H}}_\mathrm{B}[\phi(z)] -  h\phi(0) + a_1\phi^2(0)/2  
\end{equation}
where, $\tilde{\mathcal{H}}_\mathrm{B}$, is the Hamiltonian of Eq. \ref{Eq:hamB} evaluated for a z-dependent order parameter, $\phi(z)$, divided by the area of the system in the $xy$ direction. The parameters, $h$ and $a_1$, are phenomenological parameters that account for the modulation of the fluctuations at the interface. The truncation to second order in $\phi(z=0)$, restricts the analysis to weak interactions between the liquid and the electrode\cite{fredrickson1987surface}, where $a_1$ and $h$ are both order 1, accommodating non-bond interactions like van der Waals forces, and small applied potentials.  Previous observations of ordering near free interfaces in ionic liquids suggest that this approximation is sufficient\cite{jeon2012surface}.

The field $h$ is related to the chemical potential difference for positive or negative charge density at the interface. Within the assumption of weak direct surface interactions, it is expected to be linearly related to the applied potential at the electrode, $h \propto  -\Psi$\footnote{For consistency with $h$ of order 1, applied voltages should be $\Psi < k_\mathrm{B}T \ell_\mathrm{c} /z\ell_\mathrm{s}$ or about 5V for typical values of these constants at room temperature.}. For the symmetric system considered in this work, terms of zeroth order in $ \Psi$ that arise from specific chemical interactions can be neglected. In general, local interactions can give rise to terms that shift this dependence by a constant. Phenomenologically, neglecting such terms is the same as setting the zero of $h$ to the potential of zero charge. The parameter $a_1$ describes the ability of the surface to modify the local interactions between anions and cations in the electrolyte at the interface and arise due to altered packing arrangements near the weakly interacting surface. 

While it is not analytically tractable to solve for the complete partition function determined by Eq.~\ref{Eq:hamZ}, it can be approximated by neglecting fluctuations. The mean-field interfacial profile is given by
$\delta \tilde{\mathcal{H}} /\delta \bar{\phi}(\mathrm{z})  = 0$
where $\bar{\phi}(\mathrm{z})$ is the order parameter profile that minimizes the effective Hamiltonian. The resultant Euler-Lagrange equation determines the form of the profile,
\begin{equation}
\label{Eq:Euler}
a \bar{\phi}(z) - m \nabla^2  \bar{\phi}(z) + 2 \pi Q^2 \int \bar{\phi}(z')|z-z'| = 0 \, .
\end{equation}
For conditions near the bulk phase transition, the solution away from the boundary is homogeneous and the free energy is minimized by $\bar{\phi}(z)=0$. Thus, for small $h$ and $\ell_\mathrm{s}/\ell_\mathrm{c}$ close to $2$, it is sufficient to linearize Eq.~\ref{Eq:Euler} by dropping a term proportional to $\bar{\phi}^3(z)$.

\begin{figure}[t]
\begin{center}
\includegraphics[width=8cm]{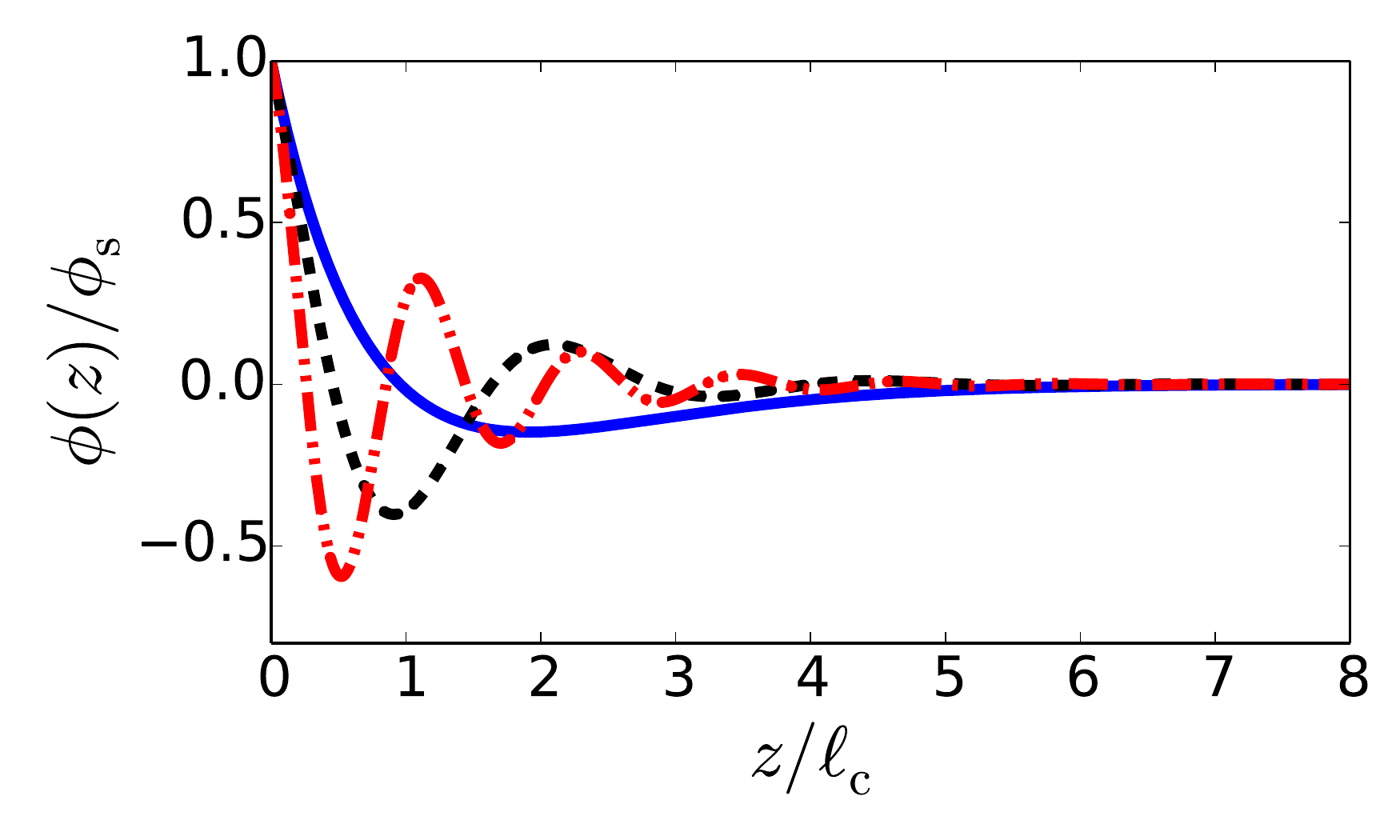}
\caption{Charge density profiles, given by Eq.~\ref{Eq:profile}, for various ratios of $\ell_\mathrm{s}/\ell_\mathrm{c}=1.5$ (blue), 0.3 (black) and 0.07 (red) and infinitesimal field, $h>0$.}
\label{Fi:2}
\end{center} 
\end{figure}

The intergro-differential equation, together with the boundary condition from the surface terms in Eq.~\ref{Eq:hamZ},
\begin{equation}
\label{Eq:BC1}
-h + a_1 \bar{\phi}(z) - m \partial_z \bar{\phi}(z)  = 0 \quad z=0 \, ,
\end{equation}
the condition that the bulk is homogenous,
\begin{equation}
\label{Eq:BC2}
\partial_z \bar{\phi}(z)  = 0 \quad z\rightarrow \infty \, ,
\end{equation}
and the constraint of electroneutrality,  $\int_\mathbf{r} \phi(\bv{r}) =0$, are sufficient to determine a unique profile for the charge density away from the electrode. The solution of this equation with these boundary conditions has the form of a damped harmonic function\cite{fredrickson1987surface},
\begin{equation}
\label{Eq:profile}
\bar{\phi}(z) = \frac{\phi_\mathrm{s}}{\cos \theta} e^{-z/\ell_\mathrm{c}} \cos \left (z q_\mathrm{s} + \theta \right )
\end{equation}
where $\tan \theta = 1/q_\mathrm{s}  \ell_\mathrm{c}$, and $\phi_\mathrm{s}$ is the value of the charge density at the surface of the electrode. The functional form of Eq.~\ref{Eq:profile} is routinely used to fit experimental data\cite{mezger2015solid}, and exhibits charge oscillations, or ``over-screening"\cite{fedorov2008ionic}, which arise from the finite size of the ions. Figure \ref{Fi:2} shows representative charge density distributions for three different values of $\ell_\mathrm{s}/\ell_\mathrm{c}$. For fixed $\ell_\mathrm{c}$ and decreasing $\ell_\mathrm{s}/\ell_\mathrm{c}$, the profile shows increased layering as a consequence of approaching the bulk phase transition.  

While the form of the charge density distribution does not change at subsequent levels of approximation, or with the incorporation of a cubic term in Eq.~\ref{Eq:hamB}, the dependence of $\phi_\mathrm{s}$ on the external field does. Within mean-field theory and for symmetric mixtures, the surface order parameter decreases smoothly as the magnitude of the external field goes to zero, with the functional form
\begin{equation}
\label{Eq:phiMF}
\phi_\mathrm{s}^{\mathrm{MF}} = \frac{h}{m\left(2/\ell_\mathrm{c} + 1/\lambda \right )}\, ,
\end{equation}
where $\lambda = a_1/m$ is the extrapolation length typically encountered in surface criticality\cite{lipowsky1983semi}. Within a self-consistent Hartree approximation,\cite{fredrickson1989kinetics,binder1997surface}, the surface order parameter is renormalized, 
$\phi_\mathrm{s}^{\mathrm{H}} = \phi_\mathrm{s}^{\mathrm{MF}} \sqrt{\Gamma}$,  
where $\Gamma$ is a strictly positive function of $a,u$ and $q_\mathrm{s}$\footnote{$\Gamma = \tilde{a}/ 3|\tilde{u} |$, where $\tilde{a} =  a + 3q_\mathrm{s}^2 u \tilde{a}^{-1/2}/2 \pi$, and $\tilde{u} =  u(1 - 3q_\mathrm{s}^2 u \tilde{a}^{-3/2}/4 \pi)/(1 + 3q_\mathrm{s}^2 u \tilde{a}^{-3/2}/4 \pi)$.}.  As the external field passes through 0, 
 $\phi_\mathrm{s}^{\mathrm{H}}$ changes discontinuously, reflecting the renormalization of the order of the phase transition\cite{brazovskii1975phase}. This discontinuous change of $\phi_\mathrm{s}^{\mathrm{H}}$ signals a first-order interfacial transition, and produces long-ranged order in a $\ell_\mathrm{c}$ thick slab parallel to the electrode, commensurate with the amplitude of the charge density wave away from the interface remaining finite for $h\rightarrow 0$. This symmetry breaking within the plane of the electrode is consistent with the onset of 2d crystallization of PF$_6^-$ ions accompanying the microphase separation observed in molecular simulations\cite{merlet2014electric}. It is also explains observations of hysteresis upon electrode charging,\cite{uysal2013structural} as nucleating domains of charge oscillations near the electrode surface will require times proportional to $\ell_\mathrm{c}q_\mathrm{s}$\cite{binder1997surface} that are large at the transition.

From Poisson's equation and the charge density in Eq. \ref{Eq:profile}, the double layer capacitance can be computed. Specifically, the applied potential is equated to the potential at the surface of the electrode by integrating Eq. \ref{Eq:profile} twice \footnote{To calculate the capacitance, the charge neutrality condition is supplanted by $\int_\mathbf{r} \phi(\bv{r}) =Q_s$, where $Q_s$ is a small accumulated charge that balances $-Q_s$ at the surface of the electrode. To first-order in $Q_s$ this alters the expression for $\theta$, and yields, $\Psi(z=0)=2 \, \ell_\mathrm{c}Q_s/(1+\ell_\mathrm{c}^2 q_\mathrm{s}^2+\ell_\mathrm{c}^2 \tilde{\phi}_s)$ as the potential at the surface relative to the bulk. The term, $\tilde{\phi}_s$ is the charge density at the surface divided by $\Psi=h$. The observed capacitance is then $C=d Q_s(\Psi)/d \Psi$.}.
The capacitance at the potential of zero charge is given by a sum of three contributions, 
\begin{equation}
\label{Eq:theta}
C/2\epsilon = \frac{1}{\ell_\mathrm{s}} + \frac{1}{\ell_\mathrm{c}} + \ell_\mathrm{c} \frac{d \phi_\mathrm{s}}{d h}\biggr \rvert_{h=0^+}  \, ,
\end{equation}
where the first two terms are expected from ideal solutions-- namely the contribution from standard Gouy-Chapman theory that is proportional to the inverse of the Debye screening length, and the Stern contribution that is proportional to the inverse of the correlation length. The last term is proportional to the surface susceptibility and is singular at an interfacial phase transition as anticipated from molecular dynamics results\cite{merlet2014electric}. Equation 9 establishes the relationship between electrolyte fluctuations parallel to the electrode and electrochemical response. Away the surface phase transition, the capacitance is found to scale as $C\propto \Delta \Psi^{-1/2}$, as found previously\cite{bazant2011double}.

While typical electrostatic calculations anticipate a bounded capacitance at finite temperature\cite{bazant2011double,skinner2010capacitance}, viewed from the perspective of classical statistical mechanics the potentially unbounded capacitance is a consequence of diverging correlation lengths encountered at a phase transition and their relationship to fluctuation and response quantities like the differential capacitance\cite{limmer2013charge}. Such collective behavior is reminiscent of charging batteries, where ion intercalation can couple to elastic modes of an electrode, resulting in discontinuous changes in accumulated charge\cite{safran1979long}. For the first-order transition found here, the capacitance is expected to diverge at the location of the surface ordering transition. For negative values of $\lambda$, the surface can order away from the bulk transition\cite{binder1972phase}. Specifically, by equating the surface free energies of the ordered and disorder phases, coexistence conditions are found at $h \rightarrow 0$ and a critical value of $\lambda=\lambda^*$, which is $\lambda^*=-4(\ell_\mathrm{c}+1)/\ell_\mathrm{c}^2$ in the mean-field case and must be solved numerically for the Hartree approximation, though $\lambda^* < -(\sqrt{2}-1)/\ell_\mathrm{c}$\cite{lipowsky1983semi}. This interfacial ordering away from the bulk transition is similar to that found in other modulated phases, such as block copolymers\cite{mansky1997interfacial} and ferroelectrics\cite{binder1972phase,bune1998two}.

\begin{figure}[t]
\begin{center}
\includegraphics[width=8.cm]{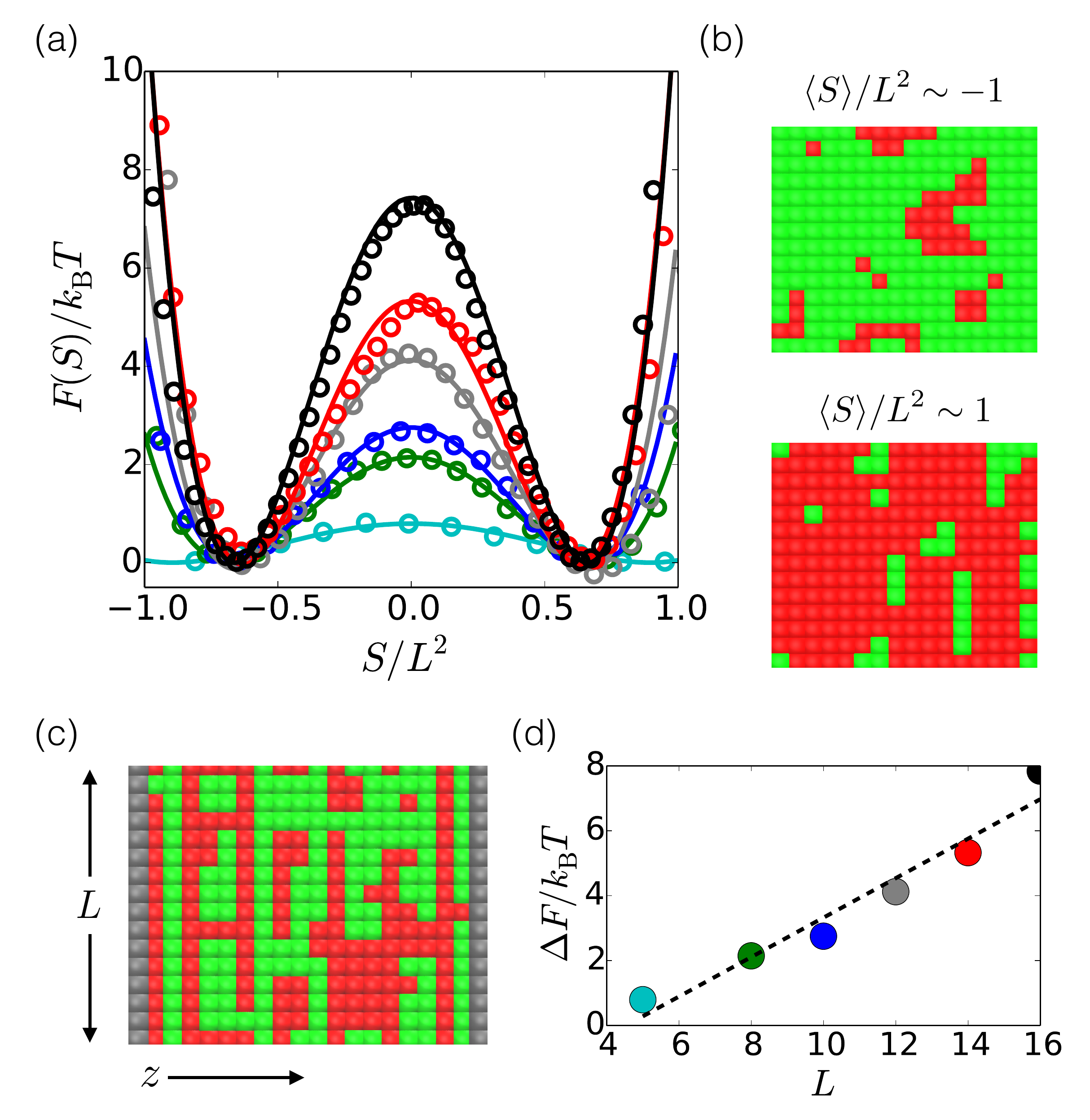}
\caption{Finite size scaling analysis for the lattice model defined in Eq.~\ref{Eq:hamCF}. a) Free energies of the accumulated charge density within the layer nearest the electrode for various electrode areas of linear size $L=5-16$. b,c) Typical configurations of the charge density, where grey locates the free boundary, $s_i=0$. d) Free energy barrier for inverting the surface charge as a function of linear electrode size. Lines are guides to the eye and errorbars are the size of the symbols.}
\label{Fi:3}
\end{center} 
\end{figure}

To check the robustness of the approximations employed above, explicit simulations of a discretized version of the system defined by the Hamiltonian in Eq.~\ref{Eq:hamB} are carried out. Specifically, I study a three-dimensional charge frustrated Ising model\cite{wu1992electrostatic} defined by
\begin{equation}
\label{Eq:hamCF}
\mathcal{H}[\{\mathbf{s}\}] = - \sum_{\langle i,j\rangle} s_i s_j + \frac{q^2}{2} \sum_{i \ne j} v(\bv{r}_\mathrm{ij}) s_i s_j 
\end{equation}
where $q$ is a reduced charge, $\{\mathbf{s}\}$ denotes the vector of Ising-like variables, $s_i = \{0, \pm1\}$, the bracket $\langle i,j \rangle$ denotes a restriction over distinct nearest neighbor pairs, $\bv{r}_{ij}$ is the distance between sites $i$ and $j$ on a three-dimensional cubic lattice and $v(r)$ is a Coulomb interaction evaluated only on those lattice sites, and asymptotically approaches $1/|\bv{r}|$ as $|\bv{r}| \rightarrow \infty$. The presence of the competing long and short ranged interactions yields bulk phase transitions in the same universality class as the Hamiltonian in Eq.~\ref{Eq:hamB} \cite{kopietz2010introduction}, only with discrete states.   

Consistent with the conditions of typical room temperature ionic liquids, I simulate a confined system in a region of the phase diagram where the bulk is disordered, but close to the first-order transition into a charge density wave phase. The bulk phase diagram for the fully occupied charge neutral case, $|s_i|=1$, has been determined from mean-field theory and explicit simulation\cite{grousson2001monte}. Simulations are run at $q=1$ and $k_\mathrm{B} T=1.2$. The simulation is embedded in an $L^2 \times L_z$ volume, with periodic boundary conditions in the $xy$ plane and noninteracting but conducting boundary conditions along the $z$ direction\footnote{The simulations are run with varying $L= L_x = L_y$, $L_z=30$ and charge swapping and displacement Monte Carlo moves, consistent with a charge neutral system. Long ranged electrostatics are handled analogously to Ref.~\onlinecite{wu1992electrostatic}. This geometry is illustrated in Fig.~\ref{Fi:3}, and $L_z$ is large enough for configurations at each surface to be uncorrelated. Conducting boundary conditions along the z direction are imposed using image charges around the plane $z=0$ and $z=L_z$, and consistent with $\Psi=0$.}. These boundary conditions explicitly generate image charges that have been neglected in the theoretical analysis but can screen ion-ion interactions. A natural order parameter that distinguishes a spontaneously polarized interfacial region from a homogeneous bulk is the amount of charge in the layer adjacent to the interface, $ \hat{S} = \sum_i s_i \, \delta (\hat{z}\cdot \mathbf{r}_i) $, which is extensive in the area of the interface, $L^2$, and approaches $\pm L^2$ in the limit that the interface is filled by positive or negative charges. This order parameter is analogous to the amplitude of the charge density wave at the electrode in the continuum limit, $\phi_\mathrm{s}$. Similarly, a divergent susceptibility of $S$ to an external field signals an interfacial phase transition and is accompanied by an singular capacitance.

Using an extension of Wang Landau sampling\cite{wang2001efficient}, I compute the relative free energy, 
\begin{equation}
F(S)/k_\mathrm{B}T = -\ln\,  \Big\langle \delta \Big(S-  \sum_i s_i \, \delta [\hat{z}\cdot \mathbf{r}_i] \Big) \big \rangle \, ,
\end{equation}
where $\langle \dots \rangle$ denotes ensemble average with fixed ions, temperature, and cell volume. The results of these calculations are shown in Fig.~\ref{Fi:3}. Each free energy curve displays two symmetric minima centered about $\pm0.7$ with a large free energy barrier in between. The height of the barrier, $\Delta F$, defined as the difference between the local maximum of free energy between $-0.5< S/L^2 < 0.5$ and its global minimum value near $S  /L^2 = \pm 0.7$ is plotted in Fig. \ref{Fi:3}(d), for various system sizes. As expected for a first-order transition in $2d$, this barrier scales linearly with $L$ and leads to a voltage dependent capacitance that diverges as $L^2$. For this symmetric solution, the symmetry breaking occurs with an infinitesimal applied field. 

{\emph{Conclusion}} I have shown how an effective field theory for the coarse-grained properties of ionic liquid metal interface can yield insight into the structural changes and responses that occur under applied voltage. By incorporating additional terms into the order parameter expansion\cite{lipowsky1983semi} or explicit image charges at the boundaries\cite{skinner2010capacitance}, the theory could be generalized to non-symmetric mixtures as well as strong direct interactions within the interface. By adopting the perspective of the capacitance as a fluctuation quantity\cite{limmer2013charge}, its relation to long-ranged correlations within the ionic liquid becomes transparent. With simple theory and numerical simulation the potential singular behavior of the capacitance at a surface phase transition is elucidated. This work highlights the importance of explicitly incorporating nonlinear behavior that arises from inter-ionic correlations.

%

\end{document}